\documentclass[12pt,preprint]{aastex}

\shorttitle{HARD X-RAY STECTRA OF CIR~X-1}
\shortauthors{Ding et al.}

\begin{document}

\title{EVOLUTION OF HARD X-RAY SPECTRA \\
ALONG THE BRANCHES IN CIR~X-1}

\author{G. Q. Ding \altaffilmark{1,2}, J. L. Qu \altaffilmark{1}, and T. P. Li \altaffilmark{1,3}}
\affil{dinggq@mail.ihep.ac.cn, qujl@mail.ihep.ac.cn, litp@mail.ihep.ac.cn}

\altaffiltext{1}{Laboratory for Particle Astrophysics, Institute of High Energy Physics, Chinese Academy of Sciences, Beijing, China}
\altaffiltext{2}{Graduate School of Chinese Academy of Sciences, Beijing, China}
\altaffiltext{3}{Department of Physics \& Center for Astrophysics, Tsinghua University, Beijing, China}

\begin{abstract}
Using the data from the PCA and HEXTE on board the RXTE satellite, we investigate 
the evolution of the 3-200 keV spectra of the peculiar low mass X-ray binary (LMXB)  
Cir~X-1 along the branches on its hardness-intensity diagram (HID) from the vertical 
horizontal branch (VHB), through the horizontal horizontal branch (HHB) and normal 
branch (NB), to the flaring branch (FB). We detect a power-law hard component in the 
spectra. It is found that the derived photon indices ($\Gamma$) of the power-law hard 
component are correlated with the position on the HID. The power-law component dominates 
the X-ray emission of Cir~X-1 in the energy band higher than $\sim 20$ keV. The fluxes 
of the power-law component are compared with those of the bremsstrahlung component in 
the spectra. A possible origin of the power-law hard component is discussed.
\end{abstract}


\section{INTRODUCTION}
Circinus~X-1 (Cir~X-1), with an orbital period of 16.6 days and a distance of 
5.5 kpc (Case et al. 1998), is thought to be a peculiar X-ray binary.
Because its rapid variability is similar to that of Cyg~X-1, its
compact object was considered to be a black hole (Toor 1977). However, the compact
object is also believed to be a neutron star because of the presence of type I
X-ray bursts (Tennant et al. 1986 a, 1986 b). The radio counterpart of Cir~X-1, 
coincident with a faint red star, displays flares with the same periodicity as
the orbital period (Whelan et al. 1977); its IR characteristics also
showed periodicity (Glass 1978); and its optical counterpart is proven to be a faint
red star (Moneti 1992). The soft X-ray (3-6 keV) flux of Cir~X-1 shows a well-defined
periodic modulation with the same period as its orbital period (Kaluzienski et
al. 1976). The large X-ray to optics luminosity ratio suggests that Cir~X-1 is 
a LMXB.

According to their fast timing properties and shapes of X-ray color-color
diagram (CCD) or the hardness-intensity diagram (HID),  LMXBs whose primary compact 
objects are neutron stars can be divided into two classes: Z source and atoll source 
(Hasinger \& van der Klis 1989). There have been different opinions about what kind 
of LMXB Cir~X-1 belongs to. From its temporal behaviors along its HID, Shirey(1998) 
believed that Cir~X-1 is a Z source; but Oosterbroek et al.(1995), after having 
studied its X-ray spectra and fast timing variations with EXOSAT ME data, suggested 
that Cir~X-1 is the only known atoll source with a very low magnetic field strength 
of less than $10^{9}$ G.

The accretion geometry of LMXBs can be inferred from studies of their hard X-ray 
spectra. Detailed studies of hard X-ray spectra of both Z and atoll sources have 
been reported (Barret et al. 2000; Di Salvo et al. 2000; D'Amico et al. 2001; Di 
Salvo et al. 2001). Using data of different high energy X-ray satellites 
(Barret et al. 1999), the X-ray and hard X-ray spectra of LMXBs, including six Z 
sources (Sco~X-1, GX~349+2, GX~340+0, Cyg~X-2, GX~5-1, and GX~17+2), have been 
analyzed on the different branches of HID or CCD.  

Using the RXTE data, Shirey et al.(1999) investigated the continuous evolutions 
of the Fourier power spectra and energy spectra of X-ray (2.5-25 keV) along 
the branches of the HID of Cir~X-1, which displays a complete ``Z'' 
track. The quasi-periodic oscillation (QPO) frequencies from 12 Hz to 30 Hz have 
been detected on the HB, 4 Hz on the NB, and no QPO on the FB. Also using RXTE 
observation data, Qu et al.(2001) studied the evolution of time lags along the 
complete HID; the time lag observed in Cir~X-1 seems to be consistent with the 
Comptonization model with a two-layer corona. 

With the ASCA observations for Cir~X-1, Brandt et al.(1996) investigated the spectral 
behavior for partial covering near zero phase of Cir~X-1, Iaria et al.(2001 a) reported 
the spectral evolution along its orbit. Using the BeppoSAX data, Iaria et al.(2001 b, 2002) 
analyzed the spectra of Cir~X-1 at the periastron and broadband spectrum at orbital phases 
close to the apoastron. However, the high energy spectra of Cir~X-1 have been poorly 
understood. In this work, using the PCA and HEXTE observations on board the RXTE, we 
investigate the evolution of broadband spectrum (3-200 keV) of Cir~X-1 along its HID from 
the VHB, through the HHB and NB, to the FB. The data selection and analysis are described 
in \S 2, and results and discussion presented in \S 3.    

\section{OBSERVATIONS AND DATA ANALYSIS}

We choose the Standard Mode 2 data of the PCA and Standard Modes (Archive) data of the 
HEXTE to perform our analysis. Because detector 2 of cluster B of the HEXTE loses its 
spectral capability and automatic gain control, only cluster A data are analyzed. The 
used observations are listed in Table 1. The ftools Version5.2 is used to perform our 
data analysis. The observations for which RXTE pointing offsets are larger than 
$0.02\rm^o$ and elevation angles less than $10\rm^o$ are discarded.  

Using the method described by Shirey et al.(1999) with soft color being defined as
the ratio of the counts in the (6.5-13.0)/(2.0-6.5) keV energy bands, we use the PCA 
data to obtain the HID of Cir~X-1, in which the soft color is plotted against the 
intensity in the 2-18.6 keV energy interval. Figure 1 shows the obtained HID with 
four branches sequentially called VHB, HHB, NB, and FB from the top left to the 
bottom right of the diagram. In the VHB, the source intensities show little variation 
with the soft color; in the HHB, the soft color remains almost constant with the source 
intensity. The soft color is positively correlated with the source intensity in the 
NB. The FB is connected to the soft end of the NB, where erratic variations are seen 
in both the hardness ratio and the source intensity.
 
We derive the hard X-ray spectrum of a certain branch according to the positions in 
the HID for each observation. The resultant hard X-ray spectrum of a branch is then 
obtained by adding all the spectra of the same branch from different OBSIDs. In order 
to enhance the signal-to-noise ratio (SNR) of the HEXTE spectra, the energy channels 
higher than 15 keV in the HEXTE spectra are rebinned. The energy channels are rebinned 
in accordance with the principle: the higher the energy channels, the more the rebinned 
energy channels.  
   
The spectra of 3-20 keV are derived from the PCA data (PCU0 only) and those of 20-200 keV 
from the HEXTE data. The dips are discarded when the the X-ray spectra are extracted. The 
HEXTE gaps and bad HEXTE observation intervals which are with zero photon count rate are 
discarded when HEXTE spectra are extracted. We use Xspec 11.2 to analyze the four broadband 
(3-200 keV) spectra. A systematic error of 1\% is added to each PCA spectrum due to the 
calibration uncertainties. The PCA X-ray spectrum and HEXTE hard X-ray spectrum of every 
branch are simultaneously fit by a combination model of a blackbody plus a line, a 
bremsstrahlung, a power-law, and an absorption edge. The fit results are shown in Figures 
2 - 5 and Table 2. From the obtained spectra, we can estimate the absorbed fluxes of the 
bremsstrahlung component of 3-200 keV, the power-law component of 20-200 keV, and the 
luminosity of 3-200 keV. The results are listed in Table 3. The F-test (Protassov et al. 2000) 
provided by Xspec 11.2 is performed to test the rationality of adding an extra power-law 
component in the spectral fitting. For each branch, the fitting $\chi^2$ and d.o.f of the 
Bblody+Line+Bremss+Edge model with and without an extra Power-law component, and the 
corresponding F-test probability are listed in Table 4.
 
\section{RESULTS AND DISCUSSION}

From the F-test probability listed in Table 4, it is shown that adding an extra  
power-law hard component is reasonable in the spectral fitting for each branch, which means
that the hard X-ray tail of Cir~X-1 is detected and its presence is not associated with a 
particular position on its HID. The derived photon indices ($\Gamma$) of the four branches 
span a range from 0.95 to 2.11 (see Table 2). The hardest index of the four branches is 
obtained when the source is on the NB, and the softest is on the VHB. 

The mass accretion rate $\dot{M}$ increases in the sequence VHB $\rightarrow$ HHB 
$\rightarrow$ NB $\rightarrow$ FB (van der Klis et al. 1996); the power law index 
continuously decreases in the sequence VHB $\rightarrow$ HHB $\rightarrow$ NB 
(see Table 2). In other words, the hard X-ray spectrum hardens from VHB, through HHB, 
to NB with the increase of the mass accretion rate $\dot{M}$. So, there may exist a 
correlation between the hardness of the hard X-ray spectrum and the mass accretion 
rate $\dot{M}$. Our results show that the hard X-ray evolution of Cir~X-1 from the 
horizontal branch (HB) to the NB is similar to that of Sco~X-1 from the HB to the 
NB (D'Amico et al. 2001).

From table 3, it is shown that hard X-ray luminosity fluxes fade in the sequence VHB 
$\rightarrow$ HHB $\rightarrow$ NB; in other words, the less the mass accretion rate 
$\dot{M}$, the larger the hard X-ray luminosity, and the softer the hard X-ray spectra. 
This behavior of the hard X-ray spectra of Cir~X-1 is similar to that shown in Sco~X-1, 
GX~17+2 (Di Salvo et al. 2000), and GX~349+2 (Di Salvo et al. 2001).   

A possible origin of the power-law hard component is the Compton up scattering of 
soft X-ray seed photons by non-thermal electrons with a power-law velocity distribution 
(Iaria et al. 2001 b). These electrons could be part of a jet in a binary system. 
Similarly, a jet could also be the origin of the power-law hard component of the detected 
hard X-ray spectra of some Z sources such as GX~5-1, GX~17+2 (Di Salvo et al. 2000), and 
GX~349+2 (Di Salvo et al. 2001). So Cir~X-1 shares the same origin of the power-law hard 
component with Z sources. It is another aspect that Cir~X-1 shares with the Z source family. 

The blackbody temperature is $\sim$ 1 keV, the blackbody component is probably emitted 
by the inner part of the accretion disk (Iaria er al. 2002). The bremsstrahlung 
seed-photon temperature of the Comptonized component is 1.2-3.1 keV. A Gaussian emission 
line with energy $\sim$ 6.5 keV and an absorption edge with energy 8.2-9.2 keV are 
detected. The iron ionization of Fe $_{\rm XXIV-XXV}$ corresponds to the emission line 
which is compatible with the absorption edge (Turner et al. 1992).  

There shows an unusual line-like structure near 10 keV on the NB and FB (see Fig. 4 and Fig. 5). 
This result is consistent with that reported by Shirey et al.(1999). Asai et al. discuss 
several mechanisms that could possibly produce a line at $\sim$ 10 keV. For example, 
emission from a heavy element such as Ni could produce the line; a line could be 
blue-shifted due to motion in a relativistic jet or rotation in the accretion disk, but 
extreme conditions would be required to boost the energy up to $\sim$ 10 keV 
(Shirey et al. 1999). That there shows an unusual line near 10 keV in the FB spectrum 
(see Fig. 5) is the reason why the reduced $\chi_{\nu}^2$ of this spectrum is as 
high as 2.54. 

It is a key goal to distinguish neutron star (NS) binaries from black hole binaries
(BHBs) in high energy astrophysics. Barret et al.(2000) suggested that only BHBs can 
have both X-ray (1-20 keV) and hard X-ray (20-200 keV) luminosities above 
$1.5\times 10^{37}$ ergs s$^{-1}$. Using 5.5 kpc (Case et al. 1998) as the distance 
to Cir~X-1, the luminosity of the power-law component in the 20-200 keV range is above 
$1.5\times 10^{37}$ ergs s$^{-1}$. The reported total X-ray flux in the 2.5-25 keV band 
of Cir~X-1 (Shirey et al. 1999) ranges over $(4.39-2.72)\times 10^{-8}$ ergs cm$^{-2}$ 
s$^{-1}$, thus the range of luminosity in the 2.5-25 keV range is 
$L_{2.5-25}^{total}=(9.85-15.89)\times 10^{37}$ ergs s$^{-1}$, and the X-ray (1-20 keV) 
luminosity must be above $1.5\times 10^{37}$ ergs s$^{-1}$. As is well known, the compact 
object of Cir~X-1 has been believed to be a neutron star because of the presence of type 
I X-ray bursts (Tennant et al, 1986 a, 1986 b), it seems that this luminosity criterion 
used to distinguish NS binaries from BHBs is not suitable to Cir~X-1.   
 
\acknowledgments
We appreciate the detailed comments and suggestions provided by the anonymous 
referee. This research has made use of data obtained through the High Energy 
Astrophysics Archive Research Center On-line Service, provided by NASA/Goddard 
Space Flight Center. We acknowledge the RXTE data teams at NASA/GSFC for their 
help. This work is subsidized by the Special Funds for Major State Basic Research 
Projects and by the National Science Foundation of China.

\clearpage

\begin{figure}
\vskip 2in
\includegraphics{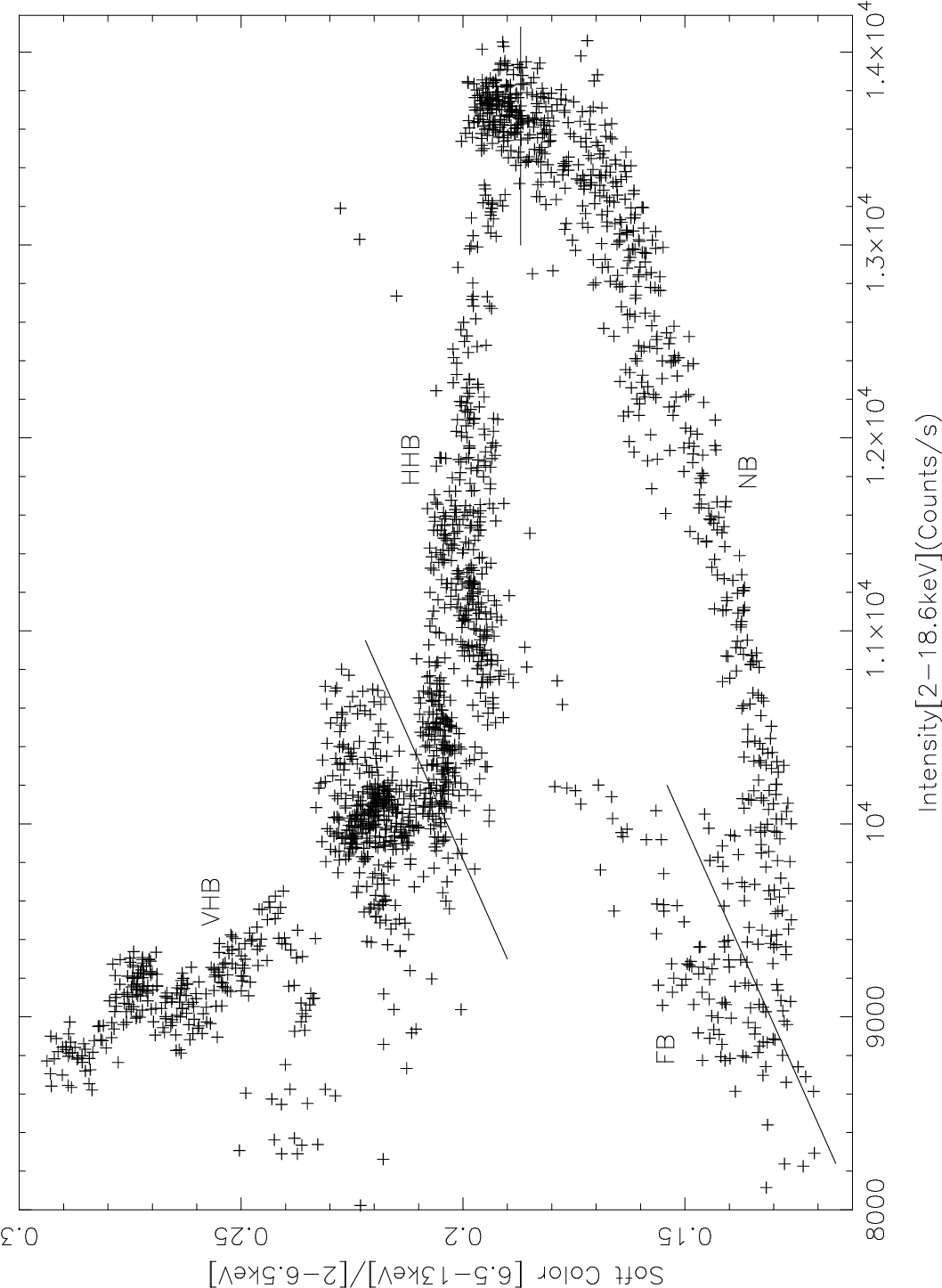}
\vskip 1.85in
\caption{The HID of Cir~X-1. Each point is from 16 s background-subtracted data 
from PCU012.}
\end{figure}

\begin{figure}
\vskip 4.2in
\includegraphics{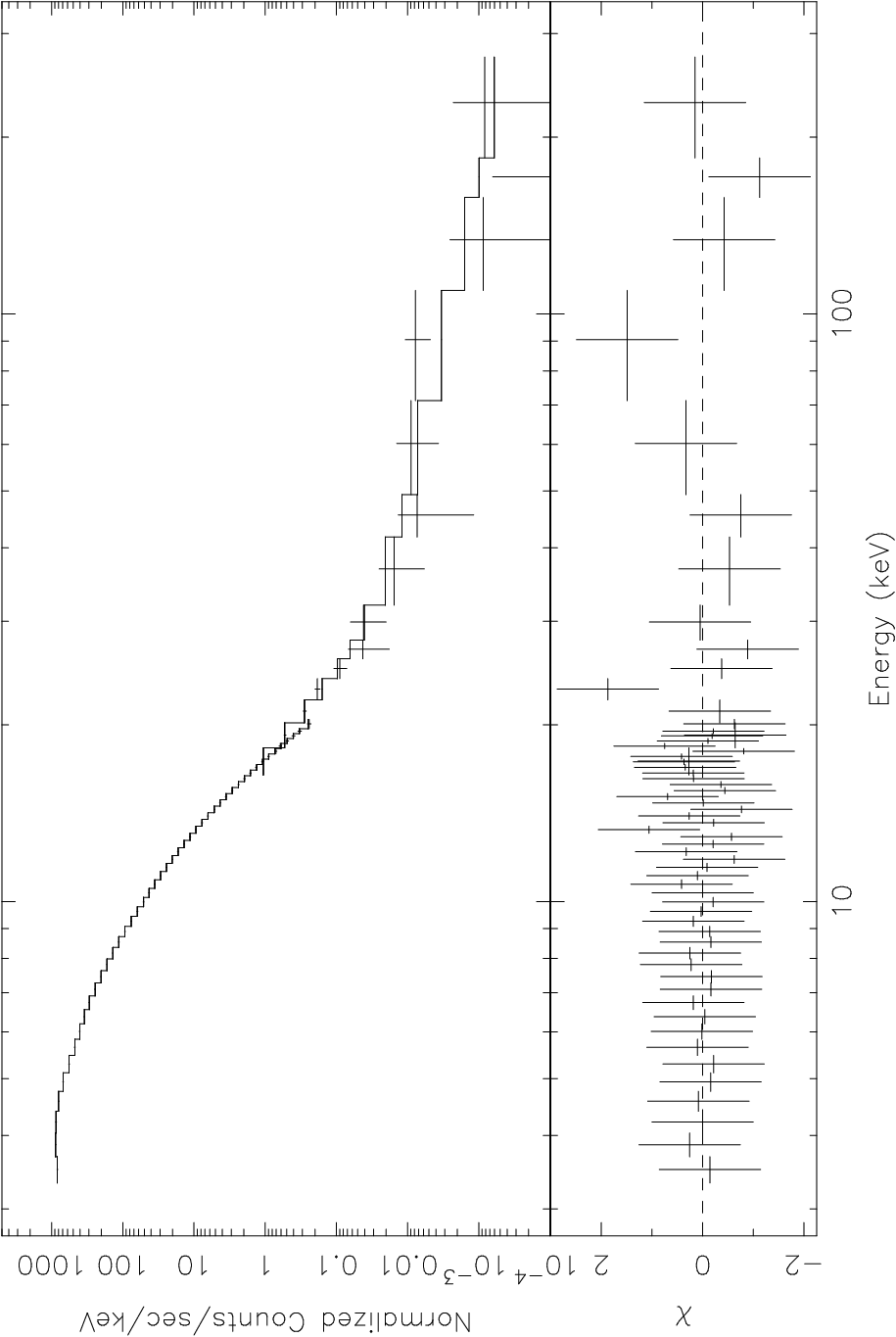}
\vskip 1.85in
\caption{The VHB Spectrum of Cir~X-1. The spectrum results from 17570 s PCA (PCU0 only) 
and 12290 s good HEXTE data. A combination model of Bbody+Line+Bremss+Power-law+Edge 
fits to the observed spectrum with  1\% systematic error being set to the PCA 
data. The reduced $\chi_{\nu}^2$ is 0.37.}
\end{figure}

\begin{figure}
\vskip 2in
\includegraphics{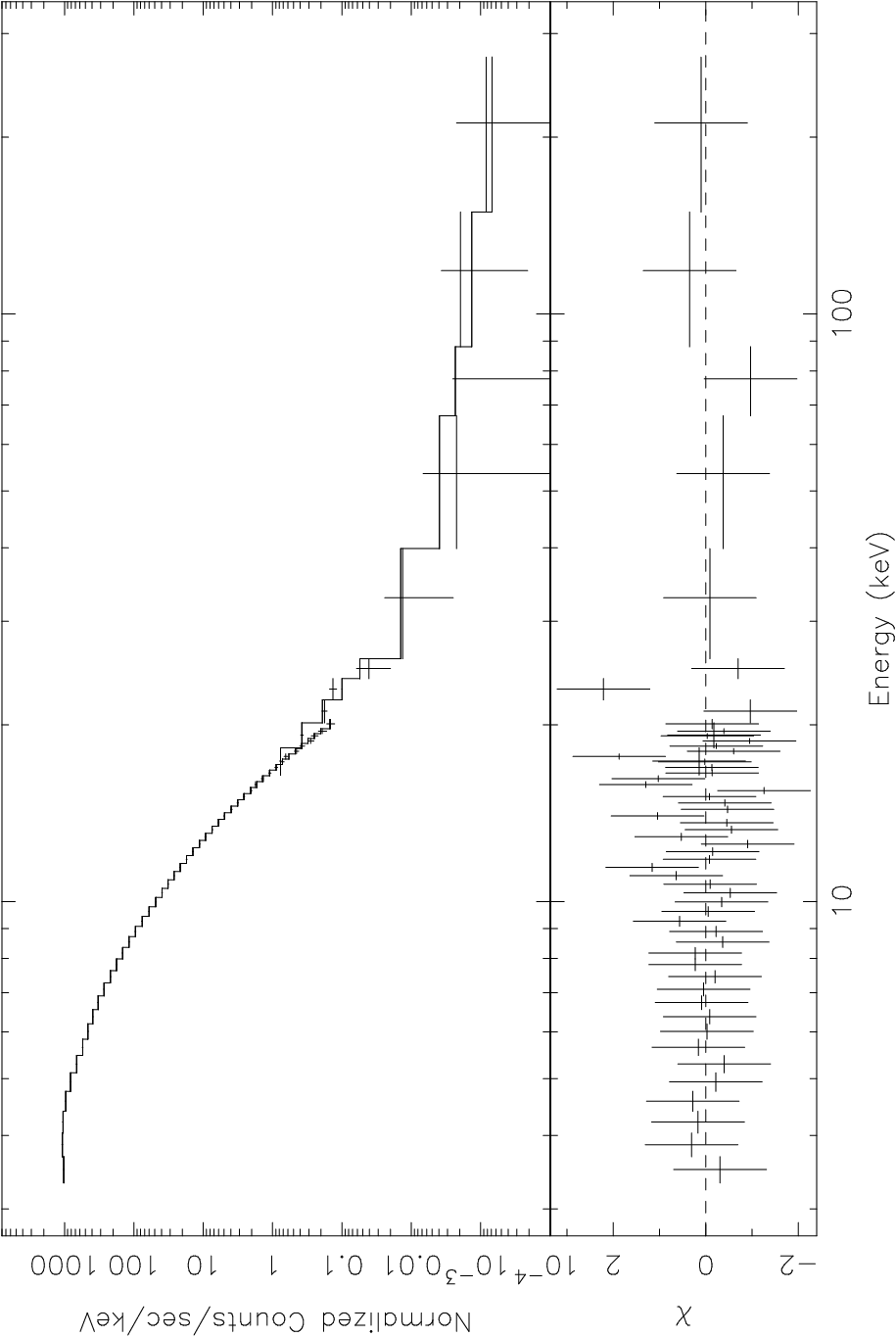}
\vskip 1.85in
\caption{The HHB Spectrum of Cir~X-1. The spectrum results from 8130 s PCA (PCU0 only) 
data and 12723 s good HEXTE data. A combination model of Bbody+Line+Bremss+Power-law+Edge 
fits to the observed spectrum with 1\% systematic error being set to the PCA 
data and the Line of E=6.50 keV being frozen. The reduced $\chi_{\nu}^2$ is 0.57.}
\end{figure}

\begin{figure}
\vskip 2in
\includegraphics{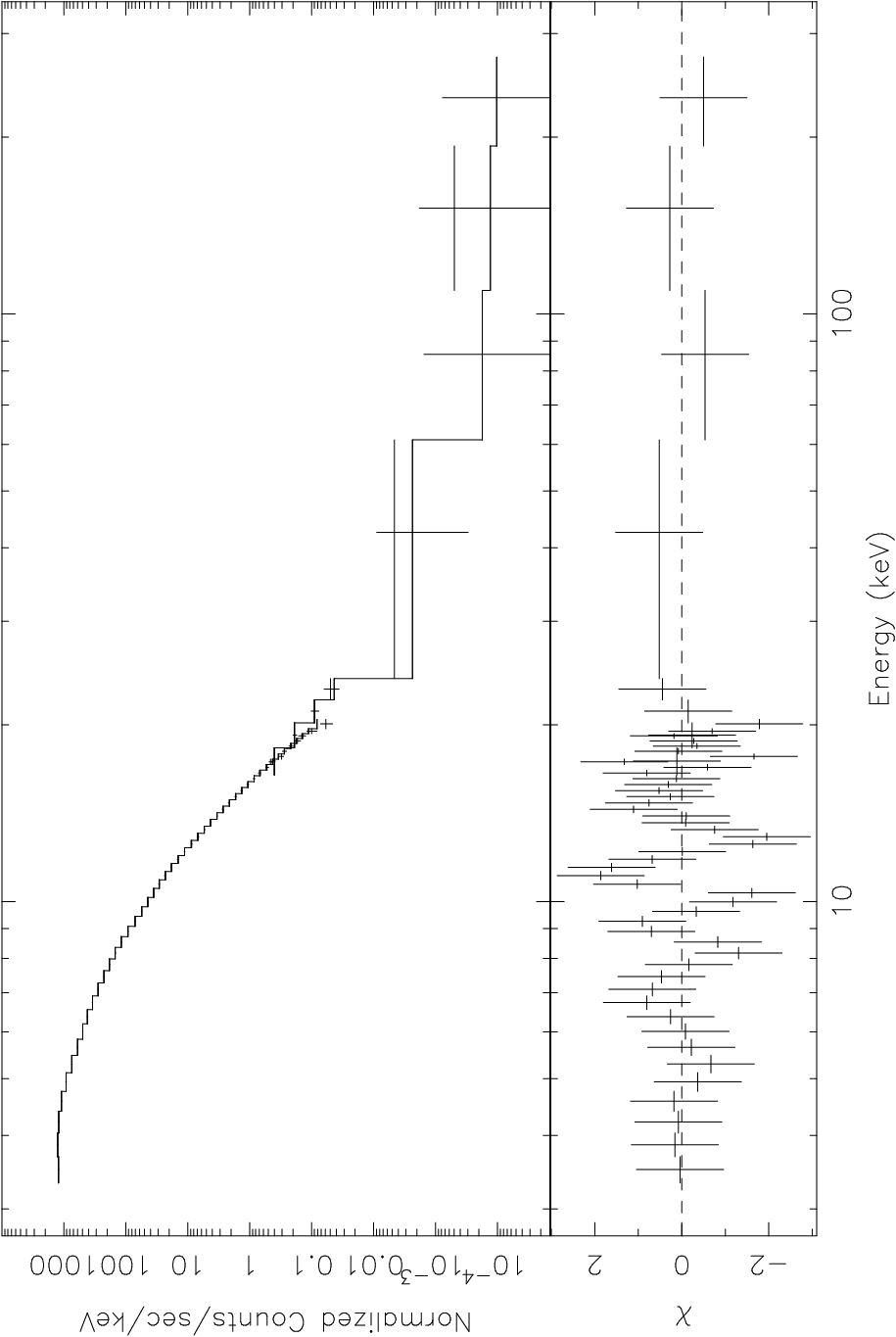}
\vskip 1.85in
\caption{The NB Spectrum of Cir~X-1. The spectrum results from 10160 s PCA (PCU0 only) 
data and 14980.0 s good HEXTE data and is fit by a combination model of Bbody+Line+Bremss+Power-law+Edge 
with 1\% systematic error being set to the PCA data and the Line of E=6.50 keV being 
frozen. The reduced $\chi_{\nu}^2$ is 0.94. There shows an unusual line near 10 keV 
in this spectrum.}
\end{figure}

\begin{figure}
\vskip 2in
\includegraphics{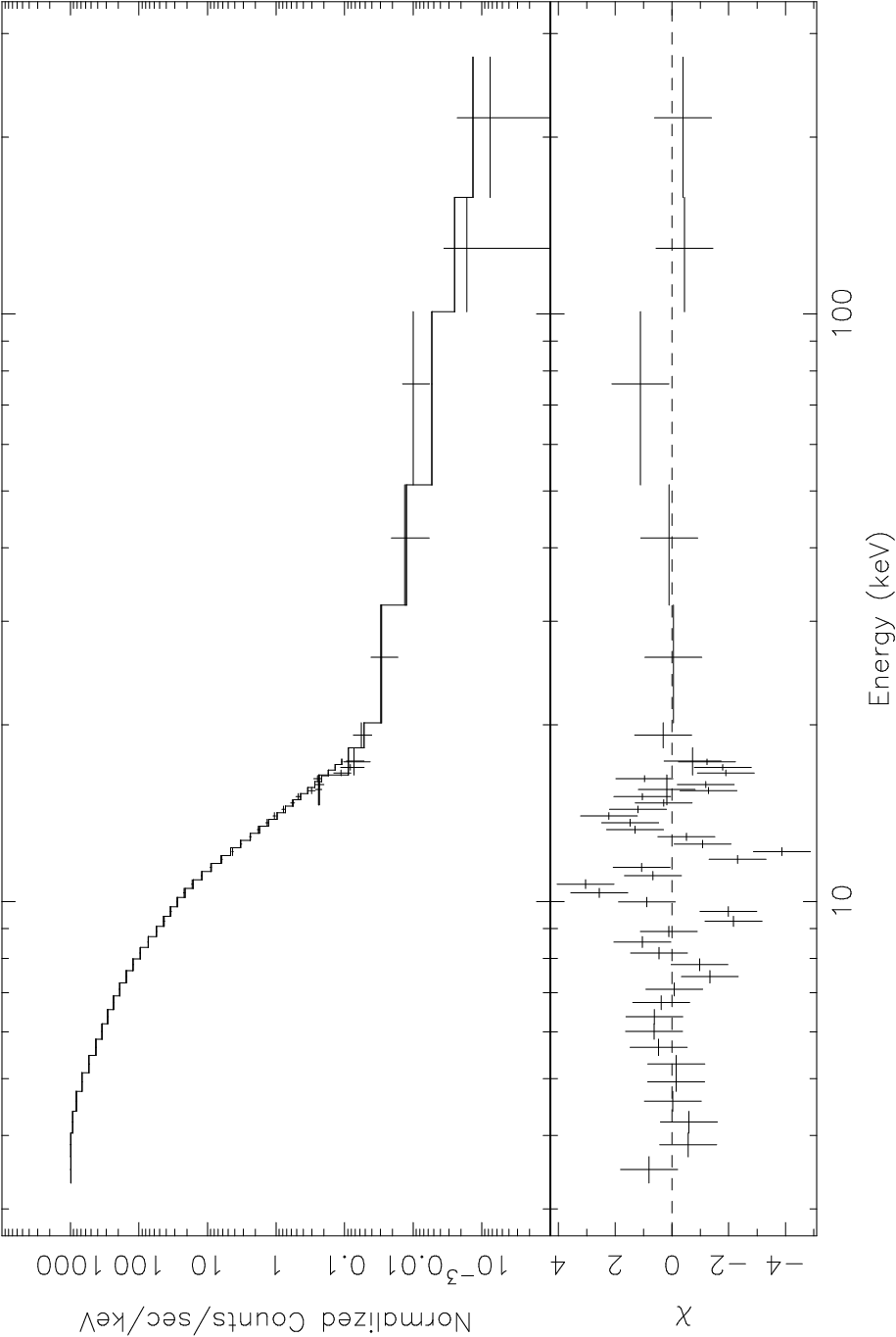}
\vskip 1.85in
\caption{ The FB Spectrum of Cir~X-1. The spectrum results from 6920 s PCA (PCU0 only) 
data and 5130.0 s good HEXTE data and is fit by a combination model of Bbody+Line+Bremss+Power-law+Edge 
with 1\% systematic error being set to the PCA data and the Line at 6.50 keV being frozen. 
The reduced $\chi_{\nu}^2$ is 2.54. That there shows an unusual line near 10 keV in this 
spectrum is the reason why the reduced $\chi_{\nu}^2$ of this spectrum is high.}
\end{figure}

\clearpage

\begin{table}
\caption{The used  RXTE observations of Cir~x-1\label{tbl-1}}
\begin{tabular}{lcccc}
\tableline\tableline
                &       &          &  PCA observation  &  HEXTE observation \\
OBSID           & MJD   & HID Pos. & Time Interval (s) & Good Time Interval (s) \\
\tableline
20094-01-02-020 & 49353 & VHB      & 17570.0           & 12290.0  \\
20094-01-02-020 & 49353 & HHB      & 8130.0            & 6265.0   \\
20094-01-02-020 & 49353 & NB       & 2352.0            & 2299.0   \\
20094-01-02-02  & 49353 & NB       & 7808.0            & 6426.0   \\
20094-01-02-02  & 49353 & FB       & 6920.0            & 4290.0   \\
20094-01-02-100 & 49353 & HHB      & 12500.0           & 6458.0   \\
20094-01-02-100 & 49353 & NB       & 4672.0            & 2665.0   \\
20094-01-02-09  & 49353 & FB       & 1100.0            & 840.0    \\
20094-01-02-09  & 49353 & NB       & 11250.0           & 3590.0   \\
\tableline
\end{tabular}
\end{table}

\begin{table}
\caption{The Fit Parameters \label{tbl-2}}
\begin{tabular}{lcccc}
\tableline\tableline
HID Pos.                              & VHB                         & HHB                       & NB                        & FB  \\
\tableline
$ N_{H}$ $(\times 10^{22}$ cm$^{-2}$) & $3.02_{-0.60}^{+0.66}$      & $2.61_{-0.75}^{+1.56}$    & $12.88_{-2.86}^{+1.85}$  & $13.25_{-2.68}^{+2.46}$  \\
\tableline
$kT_{\rm bb}$(keV)                    & $1.021_{-0.059}^{+0.052}$   & $1.026_{-0.032}^{+0.023}$ & $0.95_{-0.14}^{+0.11}$    &  $0.785_{-0.045}^{+0.049}$  \\
\tableline
$E_{\rm line}$(keV)                   & $6.48_{-0.43}^{+0.23}$      & 6.50 (frozen)             & 6.50 (frozen)             & 6.50 (frozen)  \\
\tableline
$E_{\rm edge}$(keV)                   & $9.12_{-0.25}^{+0.34}$      & $9.19_{-0.19}^{+0.21}$    & $8.820_{-0.044}^{+0.085}$ & $8.284_{-0.068}^{+0.068}$  \\
\tableline
$kT_{\rm bremss}$(keV)                & $2.92_{-0.50}^{+0.39}$      & $3.04_{-0.64}^{+0.46}$    & $2.89_{-0.44}^{+0.47}$    &  $1.215_{-0.022}^{+0.030}$  \\
\tableline
$\Gamma_{\rm power-law}$(PhoIndex)    & $2.11_{-0.69}^{+2.22} $     & $1.58_{-1.32}^{+8.42}$    & 0.95 (frozen)             &  $1.62_{-0.41}^{+0.69}$  \\
\tableline
$\chi^2~(dof)$                        &  16.08 (44)                 &  23.36 (41)               &  37.47 (40)               &  81.21 (32) \\
\tableline
\end{tabular}
\tablecomments{Blackbody+Line+Bremsstrahlung+Power-law+Edge model is used. Uncertainties are given 
at 90\% confidence level for the derived parameters of the model applied.}
\end{table}

\clearpage

\begin{table}
\caption{The Estimated Absorbed Fluxes and Luminosity \label{tbl-3}}
\begin{tabular}{lcccc}
\tableline\tableline
             &    Flux         &   Flux     & Flux            &  Luminosity  \\
             & Bremsstrahlung  & Power-law  & All Components  & All Components \\
HID Pos.     &   3-200 keV     & 20-200 keV &  3-200 keV      &  3-200 keV     \\
\tableline
VHB          &    5.43         &  10.14     &   16.10         &  5.83  \\
HHB          &    4.97         &  4.81      &   10.69         &  3.87  \\
NB           &    2.81         &  0.40      &   5.36          &  1.94  \\
FB           &    0.34         &  10.30     &   13.02         &  4.71  \\
\tableline  
\end{tabular}
\tablecomments{The fluxes are in units of $10^{-8}$ ergs cm$^{-2}$ s$^{-1}$. The luminosity is in units 
of $10^{38}$ ergs s$^{-1}$, assuming a distance of the source of 5.5 kpc (Case et al. 1988).}
\end{table}

\begin{table}
\caption{F-test of The Four Spectra\label{tbl-4}}
\begin{tabular}{lccc}
\tableline\tableline
              & Bbody+Line+Bremss   & Bbody+Line+Bremss   &   \\
              &  +Edge              &  +Power-law+Edge    &   \\
HID Pos.      & $\chi^2~ (dof)$     &  $\chi^2~ (dof) $   &  F-test Probability \\
\tableline
VHB           & 31.48  (48)         & 16.08 (44)          &  4.49 $\times 10^{-6}$ \\
HHB           & 31.64  (45)         & 23.36 (41)          &  1.27 $\times 10^{-2}$\\
NB            & 60.99  (43)         & 37.47 (40)          &  1.94 $\times 10^{-4}$ \\
FB            & 142.08 (36)         & 81.21 (32)          &  1.02 $\times 10^{-3}$ \\
\tableline
\end{tabular}
\end{table}

\end{document}